\documentclass{article}

\usepackage{arxiv}
\usepackage{graphicx}
\usepackage{algorithm}
\usepackage{algorithmic}
\usepackage{amsmath}
\usepackage{multicol}
\usepackage{multirow}
\usepackage{times} 
\usepackage[T1]{fontenc}  
\usepackage{newtxmath}  
\usepackage{dsfont}

\usepackage[utf8]{inputenc} 
\usepackage[T1]{fontenc}    
\usepackage{hyperref}       
\usepackage{url}            
\usepackage{booktabs}       
\usepackage{amsfonts}       
\usepackage{nicefrac}       
\usepackage{microtype}      
\usepackage{xcolor}         

\title{Multimodal Relational Triple Extraction with Query-based Entity Object Transformer}

\author{%
    Lei Hei\textsuperscript{\rm 1}, 
    Ning An, Tingjing Liao, 
    Qi Ma, Jiaqi Wang, 
    Feiliang Ren\textsuperscript{\rm 2}\thanks{Corresponding Author} \\
    Northeastern University, Shenyang 110169, China \\
    \textsuperscript{\rm 1}\texttt{abelcode@outlook.com}, \textsuperscript{\rm 2}\texttt{renfeiliang@cse.neu.edu.cn}
}

\begin{document}

\maketitle

\begin{abstract}
Multimodal Relation Extraction is crucial for constructing flexible and realistic knowledge graphs. Recent studies focus on extracting the relation type with entity pairs present in different modalities, such as one entity in the text and another in the image. However, existing approaches require entities and objects given beforehand, which is costly and impractical. To address the limitation, we propose a novel task, \textbf{Multimodal Entity-Object Relational Triple Extraction}, which aims to extract all triples \textit{(entity span, relation, object region)} from image-text pairs. To facilitate this study, we modified a multimodal relation extraction dataset MORE, which includes 21 relation types, to create a new dataset containing 20,264 triples, averaging 5.75 triples per image-text pair. Moreover, we propose QEOT, a query-based model with a selective attention mechanism, to dynamically explore the interaction and fusion of textual and visual information. In particular, the proposed method can simultaneously accomplish entity extraction, relation classification, and object detection with a set of queries. Our method is suitable for downstream applications and reduces error accumulation due to the pipeline-style approaches. Extensive experimental results demonstrate that our proposed method outperforms the existing baselines by 8.06\% and achieves state-of-the-art performance.
\end{abstract}

\section{Introduction}
Relation Extraction (RE) aims to identify the semantic relation types given entity pairs in the unstructured text. With the rapid development of the Internet, many studies have extended the boundaries of relation extraction using other data modalities, which benefits many multimodal downstream tasks such as visual question answering \cite{VQA_NLE, QK_VQA}, multimodal recommendation systems \cite{MMSR, MSP}, and cross-modal information retrieval \cite{CLIP, MMIR}.

Traditional relation extraction research mainly explored the ability to extract entities and predict their relations by mining textual features \cite{SPTree, GRTE, Unirel}. Recently, Zheng et al. \cite{MNRE, MNREa} introduced multimodal relation extraction (MRE) to help relation extraction models better understand relational semantics with visual features, which has emerged as a promising topic for using visual information. Further, He et al. \cite{MORE} explored the scenario in which entity pairs are present in different modalities, i.e., the head entity in the text and the tail entity in the image.

\begin{figure}[ht]
  \centering
  \includegraphics[width=0.8\linewidth]{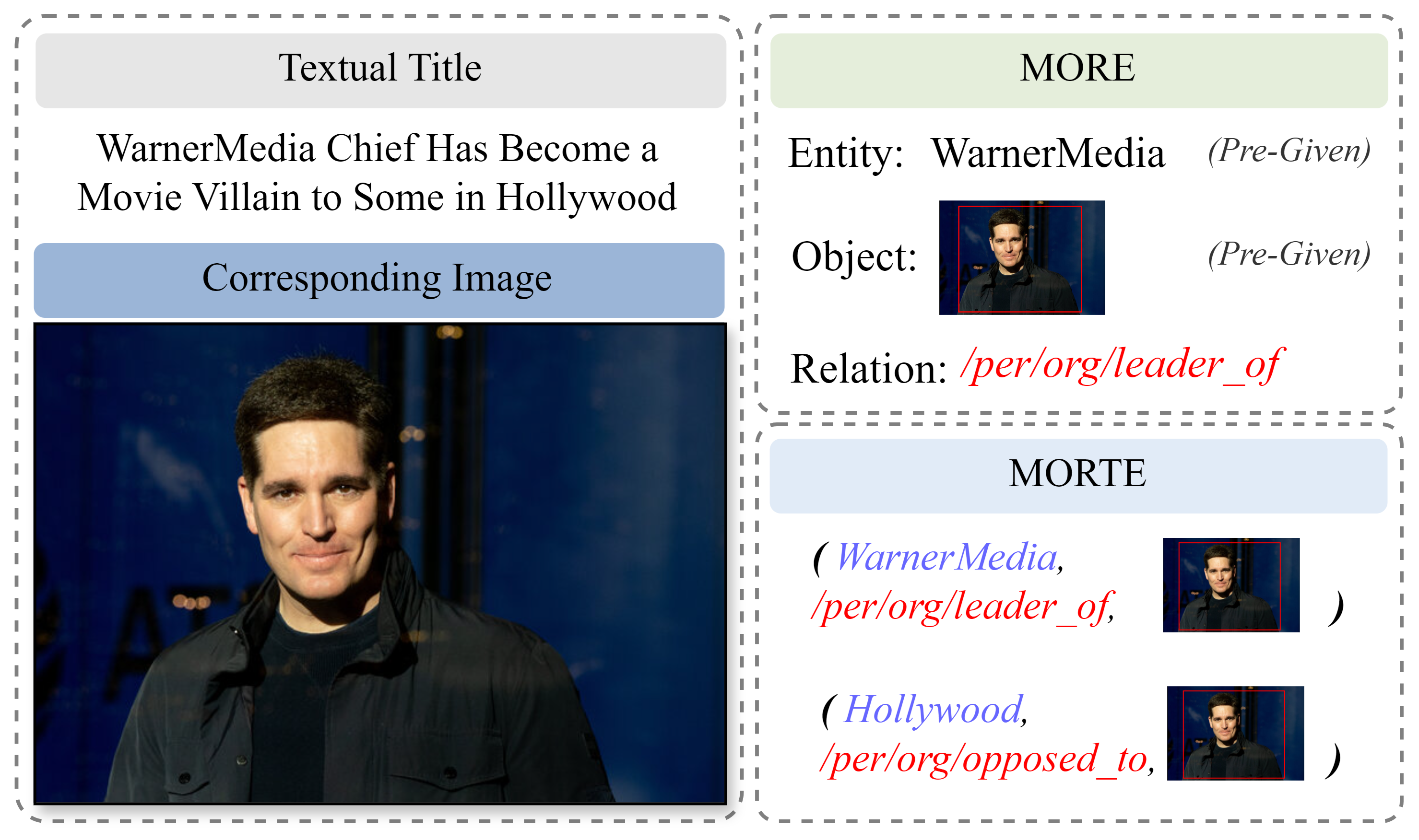}
    \caption{An example of multimodal relation extraction for entity pairs present in different modalities. Given the text and corresponding image, the MORE task only predicts the relation type between the existing entity and object. In contrast, MORTE must recognize entities, objects, and relations, plus extract all potential triples.}
    \label{fig:head}
\end{figure}

Despite the progress made by existing studies in extracting multimodal information, pressing issues still require resolution. \textbf{Firstly}, most methods leverage image information as an auxiliary cue to help with relation semantic understanding and enhance the accuracy of relation classification. However, they ignore the situation in which entities may be present in the image but not in the text, and these methods fall short when images are absent. For instance, if the news headline \textit{"the Former U.S. President Goes on Vacation in Hawaii"} lacks an image, it is challenging to determine which former president it is accurately. \textbf{Secondly}, while He et al. \cite{MORE} have explored scenarios where the head entity is in the text, and the tail entity is in the image, this approach still requires entity pairs given beforehand, yet extracting entities and objects from data is necessary for practical applications. \textbf{Thirdly}, as shown in Figure \ref{fig:head}, extracting entity pairs with semantic associations from texts and images requires considering entity extraction, relation classification, and object detection simultaneously. An intuitive way is to extract all entities using the entity extraction model, detect all objects using the object detection model separately, and match all entities and objects to predict the relation. However, this pipeline approach brings the issue of error accumulation.

Entity pairs present in different modalities and involving multiple tasks necessitate designing an end-to-end method to fulfill these requirements and minimize human involvement in entity and object extraction. As far as we know, existing research has yet to propose and tackle such a cross-modal information extraction task.

This paper introduces a new task, \textbf{Multimodal Entity-Object Relational Triple Extraction}, which aims to extract all possible triple forms \textit{(entity span, relation, object region)} from image-text pairs. To facilitate the study of this task, we provide a new dataset, \textbf{MORTE}, by modifying the dataset \textbf{MORE}, a multimodal relation extraction dataset designed for entity pairs that are not present at the same modality.\textbf{ The primary purpose} of the new task is to tackle the limitation of manually extracting entities and objects before the relation classification, further promoting the research on cross-modal information extraction. \textbf{The challenge} of the novel task is that we should not just encode visual information as auxiliary content but ensure the models can understand the visual semantics and extract objects. In addition, as shown in Figure \ref{fig:head}, an image-text pair may contain several triples since multiple entities and objects may exist in texts and images, so consequential methods must extract them comprehensively. 

In order to address the novel task, we propose an end-to-end deep learning model that employs a query-based multimodal fusion method called \textbf{QEOT} (\textbf{Q}uery-based \textbf{E}ntity-\textbf{O}bject \textbf{T}ransformer). Specifically, we use the dual-tower architecture to encode text and image information separately with the pre-trained model. We adopt selective attention and gated-fusion mechanisms to provide textual and visual embeddings with their desired features. Inspired by the object detection model DETR \cite{DETR}, we adopt a query-based approach using a set of queries to actively learn critical features in textual and visual features. Based on the multi-task learning approach \cite{MT1, MT2}, we perform classification prediction for entity span and relation type and regression prediction for object detection, thus recognizing multiple triples in the image-text pair at once. Detailed experimental results show that our model solves the task effectively, exhibits excellent performance compared to several, including the previous state-of-the-art baselines, and is well worth further investigation.

Overall, we summarize our contributions as follows:
\begin{itemize}
\item We introduce a novel task called Multimodal Entity-Object Relational Triple Extraction. This task highlights the need for more research on fusing various modalities more effectively. It also aims to meet the requirements of real-world situations, helping to create more adaptable and flexible knowledge bases.
\item Our proposed model QEOT, through a set of queries, can effectively fulfill modal fusion and simultaneously accomplish entity extraction, relation classification, and object detection tasks, providing a new end-to-end method for future research on this task.
\item We conduct comprehensive experiments and metrics to evaluate our model. Compared with recent baselines and pipeline-style models we designed, the results demonstrate the effectiveness of our approach and indicate that the novel task may be a promising research direction.
\end{itemize}

\section{Related Works}

\subsection{Text-Oriented Relation Extraction}

Relation extraction has a long history of exploration. Early relation extraction works mainly focused on feature engineering and kernel-based approaches \cite{Kernel, Kernel2} to classify relation types between entities. With the growth of deep learning, researchers are using CNN-based \cite{CNN} and RNN-based \cite{RNN, LSTM} methods to exploit text and entity features. Miwa et al. \cite{SPTree} completed an end-to-end text modeling and relation classification task using LSTM and the dependency tree approach, achieving more accurate prediction results. In order to better consider the information of the entities themselves for relation extraction, Zheng et al. \cite{Tagging} formed a joint entity and relation extraction task by tagging the beginning and end of entities and predicting their relation. However, the tagging strategy cannot solve complex cases like overlapping entities and multiple relation types. Then, CasRel \cite{Casrel} solved the tagging strategy's overlapping problem by treating the relation as a function that can map the head entities to decide which entity is the related tail entity with a fixed relation type. In recent years, some researchers treated the text as a two-dimensional table and predicted tables themselves as relations \cite{TPLinker, GRTE}, thus enabling the model to extract more complex entity and relation scenarios. UniRel \cite{Unirel} further embeds the relation semantics into the table to make the triple extraction more interpretable and robust. 

\subsection{Multimodal Relation Extraction}
Recently, several works \cite{MNRE, DGP, Syn} have focused on multimodal relation extraction tasks to extend the boundaries of natural language processing to provide sufficient semantic information by using visual information. Firstly, Zheng et al. \cite{MNREa, MNRE} adopt the scene graph generation method to create links for text and image features and construct a multimodal relation extraction dataset called MNRE,  which uses data from social media posts, thus providing additional clues for relation prediction. Subsequently, HVPNet \cite{HVPNet} designs a visual prefix-guided modal fusion mechanism that builds object-level visual information and multi-scale features, thus helping entity pairs predict more accurate relation types. IFAformer \cite{IFA} investigated the implicit alignment method between textual and visual content by creating precise, fine-grained features to combine visual features with the text and textual features with the image in the cross-attention as the key and value.

However, for previous studies, the crucial information resides in the textual content, as all entities are present within it. Visual information only augments the textual characteristics, enabling a more accurate inference of the relation between entities. In light of this, MORE \cite{MORE} asserts that it is crucial to consider the connection between text entities and image objects when dealing with real-world data, such as news. In line with this view, they devise a new task and dataset named MORE to predict the cross-modal relation type by giving a head entity (from the text) and a tail entity (from the image) at a time.

Considering the development of relation extraction, we build on MORE to further advance more realistic multimodal research by constructing triples with entity spans, relation types, and object regions, which enables the model to bridge the gap between textual and visual modalities and fully understand and explore the role of the multimodal entity-object in the information extraction. 

\section{Methodology}

\begin{figure*}[ht]
    \centering
    \includegraphics[width=\textwidth]{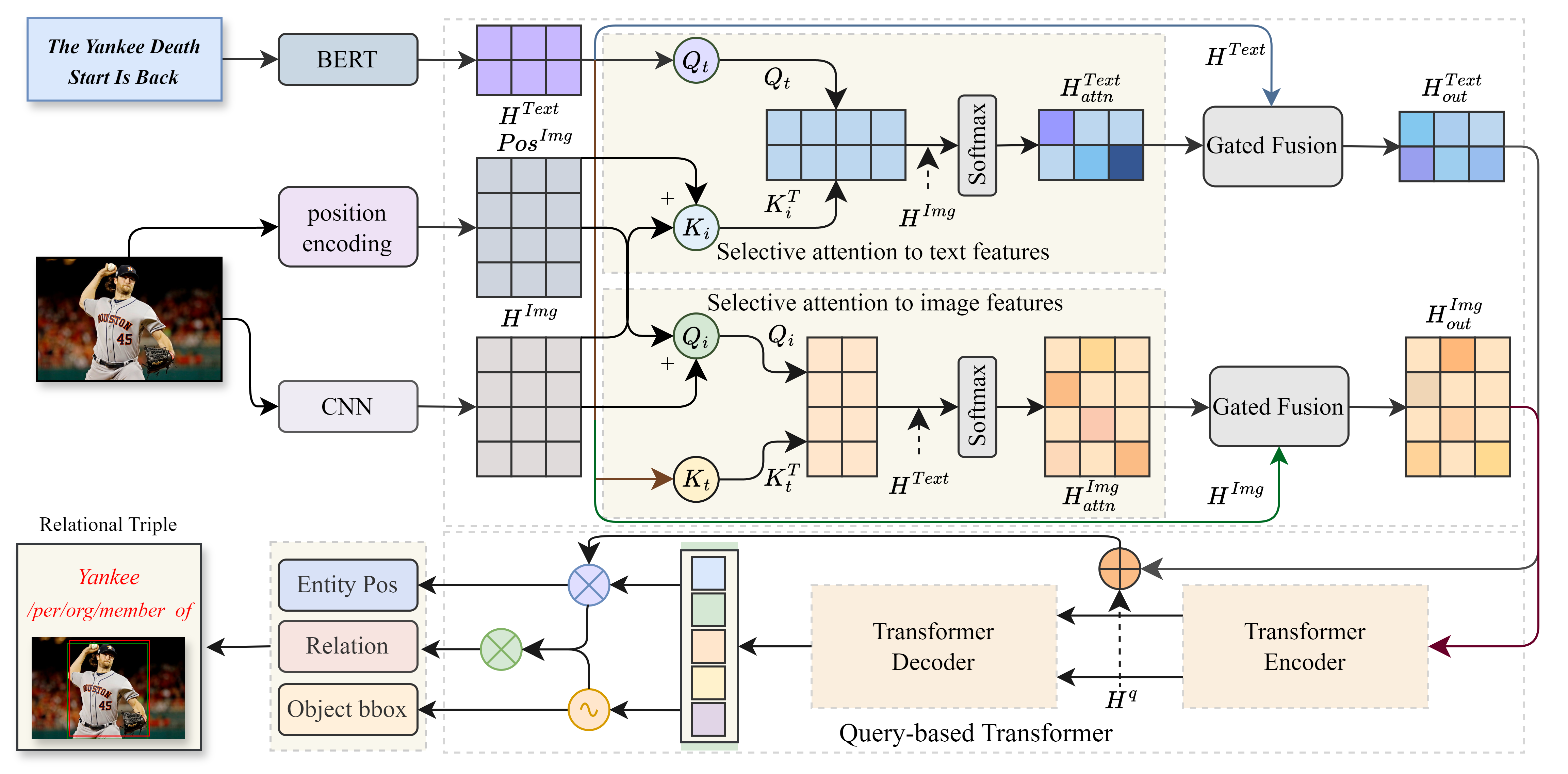}
    \caption{The overall query-based entity-object transformer architecture.}
    \label{fig:model}
\end{figure*}

This section introduces the proposed QEOT model in detail, mainly composed of three parts. Firstly, we design a selective attention mechanism to explore cross-modality representations, and the gated-fusion component helps it decide how many cross features should be left. Secondly, we introduce a query-based transformer module to learn multiple dimension features for text and image, and the set of queries can learn from the cross-modal information and other queries to select the valuable information they desire. Thirdly, the model is optimized jointly with four objectives under a comprehensive view. The overall architecture and the process are shown in Figure \ref{fig:model}.

\subsection{Task Definition}
MORE \cite{MORE} represents the second task, which predicts relations by training a function $F=(e, b, S, V) \rightarrow R$. Here, $e \in S$ is a pre-extracted entity and $b \in V$ is a pre-located visual object bounding box region. Given a sentence $S$ containing $e$ and the image $V$ containing $o$, the goal is to classify the predefined relation tag $R$.

The task of this paper, MORTE, requires extracting the entities from the text and objects from the image and predicting the relation types: given a sentence $S$ and a corresponding image $V$, the model is required to extract a set of triples $y = \{ (e, r, b)_c\}^C_{c=1}$, where $(e, r, b)_c$ represents the $c-$th triple as shown in Figure \ref{fig:head}.

\subsection{Selective Attention and Gated-fusion Mechanism}

Pre-trained language and visual models are helpful for many downstream tasks. However, the differences in structure, data, and modality of these models result in their textual and visual features not being directly used. In order to reduce the training cost of additional model parameters, we propose a selective attention mechanism and a gated fusion module to generate new embedding with multimodal features by interacting with the obtained textual and visual features. See Figure \ref{fig:model} for the architecture.

For the given sentence $S$, we used BERT \cite{BERT} to encode its textual features. For image $V$, we used ResNet \cite{ResNet}, RCNN \cite{Faster_RCNN, Mask_RCNN}, and ViT \cite{ViT} as visual encoders to fully understand the most complex part of the task, object detection. Additionally, for the model to perceive the specific position of an object in the image $V$, we add the position encoding \cite{Pos1, Pos2}. Detailed steps are as follows:

\begin{equation}
    \begin{aligned}
    H^{Text} &= TextEncoder(S) \\
    H^{V} &= VisualEncoder(V) \\
    H^{Img}, \,Pos^{Img} &= \textbf{W} H^{V}, PosEncoder(V)
    \end{aligned}
    \label{eq: eq1}
\end{equation}
where $\textbf{W}$ is a projection matrix to convert the shape of $H^V$ into $H^{Img}$ that equal with $H^{Text}$, and the $Pos^{Img}$ is the position encoding for image $V$. Note that the $VisualEncoder(\cdot{})$ can be replaced by other visual models like Faster-RCNN, ViT, etc.

After obtaining the textual and visual features from pre-trained models, we use a selective attention network with a single head to exploit the correlate tokens with image pixels (or patches) and the correlate pixels with tokens. Define the $Q_t$, $K_t$ for textual features and $Q_i$, $K_i$ for visual features, respectively, the selective attention operating the two groups of variables are defined as:

\begin{equation}
    \begin{aligned}
    H^{Text}_{attn} &= Softmax(\frac{Q_tK_t^T}{\sqrt{d_k}}H^{Img}) \\
    H^{Img}_{attn} &= Softmax(\frac{Q_iK_i^T}{\sqrt{d_k}}H^{Text})
    \end{aligned}
    \label{eq:eq2}
\end{equation}
where $d_k = d$ is the same as the dimension of $H^{Img}$ and $H^{Text}$ by using the single-head attention. $H^{Text}_{attn} \in \mathbb{R}^{L \times d}$, $H^{Img}_{attn} \in \mathbb{R}^{L \times d}$. Note the $Q$ and $K$ are defined as:
\begin{equation}
    \begin{aligned}
        Q_t = K_t &= H^{Text} \\
        Q_i = K_i &= (H^{Img} + Pos^{Img})
    \end{aligned}
    \label{eq:eq3}
\end{equation}

Finally, as a popular technique, gating mechanisms \cite{DGP, Gated} are well suited for fusing features from different sources, so we use it to help the model decide how many cross features should be left. Given the textual feature $H^{Text}_{attn}$ and visual feature $H^{Img}_{attn}$, the gate $\lambda \in [0, 1]$ and the fusion operation are defined as:

\begin{equation}
    \begin{aligned}
        \lambda &= \sigma(\textbf{A} H^{Text}_{attn} + \textbf{B} H^{Text}) \\
        H^{Text}_{out} &= (1 - \lambda) \cdot H^{Text} + \lambda \cdot H^{Text}_{attn}
    \end{aligned}
    \label{eq:eq4}
\end{equation}
where $\sigma$ is the \textit{sigmoid} activation function, $\textbf{A}$ and $\textbf{B}$ are trainable matrix. By replace the $H^{Text}$ and $H^{Text}_{attn}$ with $H^{Img}$ and $H^{Img}_{attn}$ respectively, the $H^{Img}_{out}$ will be produced. Note $H^{Text}_{out}$ and $H^{Img}_{out} \in \mathbb{R}^{L \times d}$.

\subsection{Query-based Transformer}
Since an image-text pair may contain multiple triples, we hardly directly convert it to a multi-class classification task. We adopt the query-based Transformer structure \cite{Transformer, DETR} to simplify this task and ensure the model operates end-to-end. 

Specifically, we randomly initialize a set of queries, as shown in Figure \ref{fig:model}, let the multimodal features obtained earlier go through the transformer encoder, and then the queries interact with the relevant features in the transformer decoder to allow each query to extract its desired features. The transformer encoder specializes in multimodal sequences. We take feature $H^{Img}_{out}$, which contains more information about the original image, as the input to the encoder, and each layer of the encoder is a standard multi-head self-attention (MSA) module and feed-forward network (FFN). The dimensions are identical for input and output sequences.

The attention and feed-forward network modules in the transformer decoder mirror the standard modules, albeit with differences in the inputs they process and the sequence flow. The inputs for each decoder layer consist of multimodal sequences derived from the transformer encoder, along with defined queries. These queries undergo multi-head self-attention before cross-attention operations with the multimodal sequences. The reason for designing such a process is to allow the queries to discern the features obtained by other queries and subsequently determine the features they will extract from the multimodal sequences.

In addition to simplifying the difficulty of multiple triples, the features learned from different queries facilitate the prediction of entity span, relation type, and object region. After getting the output queries $H^{q} \in \mathbb{R}^{Q \times d}$, which $H^{q} = QueryTransformer(H^{Img}_{out})$, from the query-based transformer ($Q$ is the number of queries, $d$ is the hidden size), we will fulfill the task in a multi-task learning way. Since entities are more inclined to textual features for the entity extraction task, to avoid the query losing features during the learning process, we first do the residual operation before predicting the entities' position $E_{pos}$ as follows:

\begin{equation}
    \begin{aligned}
        H^{ent} &= H^{q}[L;1] + H^{Text}_{out}[Q;0] \\
        E_{pos} &= \textit{ReLU}\,(H^{ent} W_{pos} + b_{pos})
    \end{aligned}
    \label{eq:eq5}
\end{equation}
where $H^{q}[L;1] \in \mathbb{R}^{Q \times L \times d}$ denote to repeat $H^{q}$ for $L$ times at the first dimension, similarly, $H^{Text}_{out}[Q;0] \in \mathbb{R}^{Q\times L \times d}$. The $H^{ent}$ is a residual variable, and $L$ denote the sequence length. $W_{pos}$ and $b_{pos}$ are trainable variables to classify the entity span, \textit{ReLU} \cite{Relu} is an activation function. Additionally, we defined the $H^{cross} \in \mathbb{R}^{2d}$ as the concated mean tensors of $H^{Text}_{out}$ and $H^{Img}_{out}$.

We make a regression prediction directly on $H^{q}$. The bounding box comprises four points for the object region. Hence, we must use the sigmoid function to keep it in [0, 1]. When predicting the relation type, we must consider entity and object information together. More importantly, when classifying the relation type, we need an extra empty category $\varnothing$ for \textit{no relation}, which means there are $R+1$ types in total. We do not output the $\varnothing$ type when we evaluate the performance to avoid generating $Q$ triples for each image-text pair, which would result in a drop in precision. The bounding boxes $B_{box}$ and relations $R_{rel}$ are defined as:

\begin{equation}
    \begin{aligned}
        H^{rel} &= H^{cross}[Q;0] + H^{q} \\
        R_{rel} &= ReLU (H^{rel} W_{rel} + b_{rel}) \\
        B_{box} &= \sigma (\textit{ReLU}\,(H^{q} W_{obj} + b_{obj})) \\
    \end{aligned}
    \label{eq:eq6}
\end{equation}
where $H^{cross}[Q;0] \in \mathbb{R}^{Q\times 2d}$ and $H^q \in \mathbb{R}^{Q \times d}$ do the broadcast and residual operation, $\sigma$ is the sigmoid activation function, $W_{rel}$ and $b_{rel}$ are trainable variables for relation prediction, $W_{obj}$ and $b_{obj}$ are trainable variables for object region detection.

\subsection{Joint Optimization}
Unlike single-task learning, when the model needs to output many different predictions at the same time, we have to adopt the joint optimization method to improve the model's performance in an end-to-end way. For the multimodal entity-object relational triple extraction task, we must consider the joint optimization of three sub-tasks of entity extraction, relation classification, and object detection. Essentially, entity extraction is a probabilistic prediction of the position for a given text sequence, relation classification is a typical multi-class classification task, and object detection requires regression prediction of the four region points. 

Further, since multiple outputs are present at a time and bounding boxes are not discrete classes, we must find a way to match the predicted and ground truth pairwise. Based on the above analysis, let us denote by $y$ the ground truth set of objects, and $\hat{y} = \{\hat{y}_{i=1}^Q\}$ the set of Q predictions. The permutation of Q elements $\sigma \in \Delta_Q$ with the lowest cost is computed by:

\begin{equation}
    \begin{aligned}
        \sigma = \mathop{\arg\min}\limits_{\sigma \in \Delta_{Q}} \sum^{Q}_{i} \mathcal{L}_{match}(y_i, \hat{y}_{\sigma(i)})
    \end{aligned}
    \label{eq:eq7}
\end{equation}
where $\mathcal{L}_{match}(y_i, \hat{y}_{\sigma(i)})$ is the matching cost between $y_i$ and a prediction with index $\sigma(i)$. The optimal assignment is computed with the Hungarian algorithm \cite{BM}, which is designed for the bipartite match problem from the graph theory.

The matching cost is calculated by considering entity and relation predictions and the similarity of predicted and ground truth boxes. Each element $i$ of the ground truth can be seen as $y_i = (e_i, r_i, b_i)$ where $e_i$, $r_i$ and $b_i$ are the target entity, relation (which $r_i \neq \varnothing$) and object respectively. The relational triples' prediction can be seen as $\hat{y}_{\sigma(i)} = (\hat{p}^{e}_{\sigma(i)}, \hat{p}^{r}_{\sigma(i)}, \hat{b}_{\sigma(i)})$, where the $\hat{p}^{e}_{\sigma(i)}$ and $\hat{p}^{r}_{\sigma(i)}$ are the probability of entity $e_i$ as $\hat{p}_{\sigma(i)}(e_i)$ and relation $r_i$ as $\hat{p}_{\sigma(i)}(r_i)$. The $\hat{b}_{\sigma(i)}$ is the predicted bounding box. With these notations, we define $\mathcal{L}_{match}(y_i, \hat{y}_{\sigma(i)})$ as:

\begin{equation}
    \fontsize{7.5pt}{12pt}\selectfont
    \begin{aligned}
     \mathcal{L}_{er} &= -\mathds{1}_{\{r_i \neq \varnothing \}}\hat{p}^{e}_{\sigma(i)} - \mathds{1}_{\{r_i \neq \varnothing \}}\hat{p}^{r}_{\sigma(i)} \\
     \mathcal{L}_{bbox}(b_i, \hat{b}_{\sigma(i)}) &= \lambda_{giou}\mathcal{L}_{giou}(b_i, \hat{b}_{\sigma(i)}) + \lambda_{L1}\Vert{b_i - \hat{b}_{\sigma(i)}}\Vert_1 \\
     \mathcal{L}_{match}(y_i, \hat{y}_{\sigma(i)}) &= \mathcal{L}_{er} + \mathds{1}_{\{r_i \neq \varnothing \}} \mathcal{L}_{bbox}(b_i, \hat{b}_{\sigma(i)}) \\
    \end{aligned}
    \label{eq:eq8}
\end{equation}
where $\lambda_{giou} \in \mathbb{R}$ and $\lambda_{L1} \in \mathbb{R}$ are hyperparameters, and $\mathcal{L}_{giou}$ is the generalized IoU loss \cite{GIoU}. It is important to note that these operations only compute the optimal assignment and thus do not involve gradient back-propagation. In our final computation, entity extraction and relation classification will use the cross-entropy loss, denoted $\mathcal{L}_{ent}$ and $\mathcal{L}_{rel}$, respectively. The \textit{L}1 loss and generalized IoU loss still compute the object.

Our final jointly optimized loss function is:
\begin{equation}
    \fontsize{8.5pt}{12pt}\selectfont
    \begin{aligned}
		\mathcal{L} = \lambda_{ent}\mathcal{L}_{ent} + \lambda_{rel}\mathcal{L}_{rel} + \lambda_{L1}\Vert{b_i - \hat{b}_{\sigma(i)}}\Vert_1 + \lambda_{giou}\mathcal{L}_{giou}
	\end{aligned}
	\label{eq:eq9}
\end{equation}
where $\lambda_{*}$ represents hyperparameters used to control the effects of different loss functions.

\section{Experiments}

\subsection{Experimental Settings}

\renewcommand{\arraystretch}{1}
\begin{table*}[ht]
    \centering
    \small
\resizebox{\textwidth}{!}{%
    \begin{tabular}{c | c c c  c c c c c} 
        \toprule
            \multirow{2}{*}{Method}& 
            \multicolumn{3}{c}{Entity-Object Triple} & \multicolumn{3}{c}{Entity-Relation Pair} & \multirow{2}{*}{Rel Acc.} & \multirow{2}{*}{Ent Acc.} \\
            \cmidrule(lr){2-4} \cmidrule(lr){5-7} &\textit{Precision}&\textit{Recall}&\textit{F1 Score}&\textit{Precision}&\textit{Recall}&\textit{F1 Score}\\
        \midrule 
            IFAformer* & 37.86 & 13.07 & 19.43 & 50.38 & 29.29 & 37.04 & 41.73 & 53.29 \\
            HVPNeT* & 36.28 & 13.73 & 19.92 & 53.07 & 30.86 & 39.03 & 40.87 & 57.87 \\
            MOREformer* & 33.03 & 9.7 & 15.00 & 49.23 & 28.36 & 35.99 & 40.04 & 58.01 \\
            VisualBERT* & 35.31 & 14.87 & 20.93 & 57.85 & 32.87 & 41.92 & 43.84 & 61.21 \\
            \hline
            Pipeline-R50 &  41.34 & 33.43 & 36.97 & 52.11 & 48.94 & 50.48 & 59.91 & 78.40 \\
            Pipeline-R101 & 40.90 & 27.71 & 33.04 & 51.05 & 43.97 & 47.25 & 51.45 & 77.99 \\
            Pipeline-RCNN & 43.85 & 39.55 & 41.59 & 56.52 & 51.89 & 54.11 & 61.61 & 74.20 \\
            \hline
            QEOT & \textbf{48.65} & \textbf{50.70} & \textbf{49.65} & \textbf{64.20} & \textbf{61.38} & \textbf{62.76} & \textbf{68.31} & \textbf{87.36} \\
            Improve-R & \textit{+4.80} & \textit{+11.15} & \textit{+8.06} & \textit{+7.68} & \textit{+9.49} & \textit{+8.65} & \textit{+6.70} & \textit{+13.16} \\
            Improve-V & \textit{+13.34} & \textit{+35.83} & \textit{+28.73} & \textit{+6.35} & \textit{+28.51} & \textit{+20.84} & \textit{+24.47} & \textit{+26.15} \\
            \bottomrule
  
    \end{tabular}
}
    \caption{Overall performance comparison of all models on MORTE dataset. Rel Acc. denotes the relation accuracy and Ent Acc. denotes the entity accuracy. The bold values indicate the best performance. Improve-R denotes the improved value of our model compared to the Pipeline-RCNN model, and Improve-V denotes the improved value compared to VisualBERT.}
    \label{tab:table2}
\end{table*}

\subsubsection{Dataset}
To facilitate the study of the Multimodel Entity-Object Relational Triple Extraction task, we modify the MORE \cite{MORE} to obtain the MORTE dataset. We keep the same with MORE as much as possible. Notably, MORE contains 20264 train samples because each of their samples contains the prediction of only one type of relation. In contrast, we use a unique combination of each image-text pair, which makes the number of 3559 samples and the total number of our relation triples remain 20264.

\subsubsection{Baselines}
We compare the proposed QEOT with several multimodal relation extraction models to evaluate its performance. Baselines include IFAformer, HVPNeT, MOREformer, and VisualBERT. Since these models primarily address relation extraction tasks with pre-given entity pairs, we adapted them as end-to-end models with entity extraction and object detection to fit the proposed task.

Due to the complexity of the proposed task, it is unfair to compare these multimodal relation extraction models dealing with a single task, so we also designed three pipeline-style baselines: Pipeline-R50 combine the ResNet-50 \cite{ResNet} with the BERT encoder to deal with visual and textual features. We train the entity extraction and object detection separately for the models and add a relational classification module after training them. Pipeline-R101 is replacing the ResNet-50 backbone with the ResNet-101. Pipeline-RCNN is similar to other pipeline models, and we replace the ResNet-* backbone with Faster-RCNN \cite{Faster_RCNN} to evaluate the object detection backbone.

\subsubsection{Evaluation Protocol}
This paper focuses on evaluating the multimodal entity-object relational triple extraction task. Since a single image-text pair may generate multiple triples, we established three types of metrics to assess model performance. (1) Entity and relation accuracy evaluate the model's ability to understand entity spans or relation semantics in a single dimension. (2) F1 score, precision, and recall (abbreviated as FPR) are used to assess entity-relation pairing. (3) Since the objects in this task do not have fixed categories but instead generate semantic associations with the head entity, making mAP evaluation unsuitable. To address this, we developed an algorithm similar to Equation \ref{eq:eq8} to optimally match object positions based on the entity-relation pairs and ground truth, and to determine the FPR evaluation for all triples. The detailed procedure is described in the Appendix.

\subsection{Main Results}
We conducted a comprehensive experiment based on the proposed MORTE dataset with FPR metrics. The comparison results of our QEOT method with other baseline models are shown in Table \ref{tab:table2}. Based on the experimental results, we have the following analysis:

\textbf{Firstly}, our proposed QEOT method outperforms other models, VisualBERT, by 28.73\% and Pipeline-RCNN, by 8.06\% (triple f1 score). The experimental results verify the effectiveness of our model.

\textbf{Secondly}, our models perform better than other multimodal relation extraction models for two reasons: (1) These models are constructed for single-task learning. They cannot handle multi-task learning and multiple outputs. (2) These models need entity pairs and extra information given beforehand to help them understand relation semantics, yet the proposed task only provided an image-text pair. MOREformer is a SOTA model that proves a significant performance drop without insufficient extra information (such as object captions and depth features).

\textbf{Thirdly}, the \textit{Pipeline-$*$} models are better than multimodal relation extraction models because they extract all possible entities and objects sequentially and then enumerate them for relation classification without extra information. However, they are still much worse than our model, as the pipeline approach brings an error accumulation issue. Moreover, these models cannot solve the problem since they have no gradient update. Counterintuitively, Pipeline-R50 outperforms Pipeline-R101. We attribute this to the proposed dataset's smaller size, making it more challenging to transfer the task when fine-tuning larger pre-trained models.

\textbf{Lastly}, observing each metric, our model is more accurate at predicting entities and relations than other models. Because the entities are represented as head entities in the proposed task, the relation must account for cross-modal semantic correlation, unlike other models that lack feedback on identifying the head entity and tail object.


\begin{table*}[ht]
    \centering
\resizebox{\textwidth}{!}{%
    \begin{tabular}{c | c c c c c c c c}
    \toprule
         & Triple Prec. & Triple Rec. & Triple F1. & Pair Prec. & Pair Rec. & Pair F1. & Relation Acc. & Entity Acc. \\
    \hline
    QEOT & \textbf{48.65} & \textbf{50.70} & \textbf{49.65} & \textbf{64.20} & \textbf{61.38} & \textbf{62.76} & \textbf{68.31} & \textbf{87.36} \\
    \textit{r/p r50} & 36.53 & 20.32 & 26.11 & 48.76 & 40.31 & 44.13 & 50.00 & 74.30 \\
    \textit{w/o gated} & 46.81 & 43.74 & 45.22 & 60.64 & 55.11 & 57.74 & 63.79 & 81.94 \\
    \textit{w/o select} & 41.36 & 45.72 & 43.43 & 63.60 & 60.60 & 62.07 & 66.95 & 83.40 \\
    \textit{w/o query} & 38.70 & 29.09 & 33.21 & 65.80 & 44.62 & 53.18 & 59.12 & 69.44 \\
    \bottomrule
    \end{tabular}
}
    \caption{Ablation study on critical components. The order of metrics is the same as in Table \ref{tab:table2}, Triple denotes the Entity-Object Triple, Pair denotes the Entity-Relation Pair.}
    \label{tab:table3}
\end{table*}

The results show that our method's modal fusion and multi-task learning effectively solve the proposed task. Our end-to-end approach can avoids the error accumulation problem. Most importantly, experiments demonstrate that previous state-of-the-art models can only solve the proposed task with sufficient additional information. These analyses suggest that the task proposed in this paper proves the need for further multimodal semantic understanding research.

\begin{figure}[ht]
    \centering
    \includegraphics[width=\textwidth]{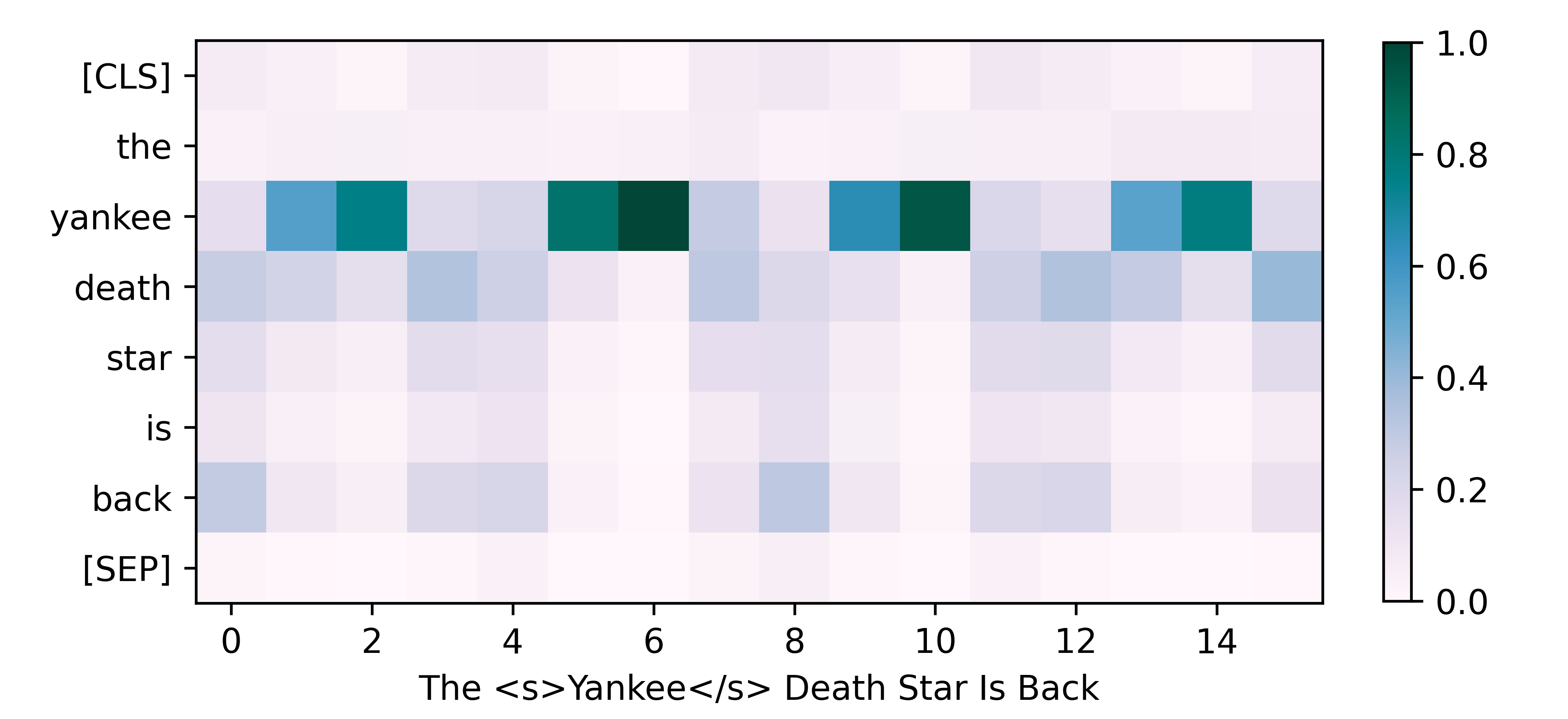}
    \caption{Visualization of selective attention of image to text.}
    \label{fig:text}
\end{figure}

\begin{figure}[ht]
    \centering
    \includegraphics[width=\textwidth]{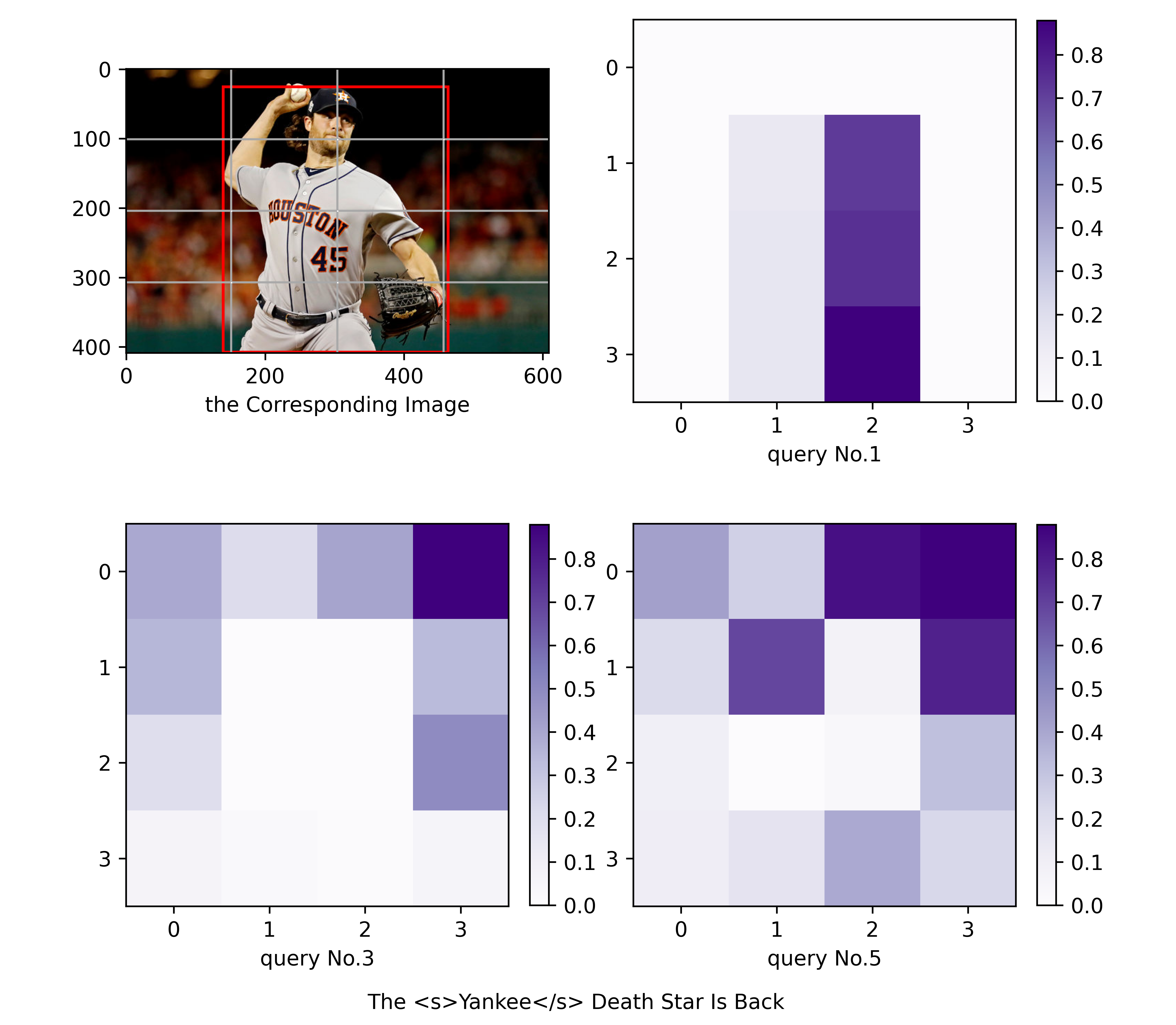}
    \caption{Visualization of cross-attention of queries to image.}
    \label{fig:query}
\end{figure}

\subsection{Ablation Study}

We designed the following four ablation experiments to validate the impact and role of each key module of our proposed method in the model. (1) We replace the Faster-RCNN with ResNet-50 as our image backbone to validate the role of the visual encoder, denoted as \textit{r/p r50}. (2) We remove the gated-fusion module and do not use the dynamic selected cross-model features to validate the fusion mechanism, denoted as \textit{w/o gated}. (3) We remove the entire selective attention component to verify the performance when text and image features directly predict triples without interacting, denoted as \textit{w/o select}. (4) We remove the crucial query-based transformer module to judge its ability to cross-modal features and the ability of multi-task learning, denoted as \textit{w/o query}. The experimental results are shown in Table \ref{tab:table3}. According to the results, we can observe the following conclusions:

\begin{itemize}
    \item When we use ResNet-50 as the backbone, all metrics degrade, indicating that the object detection models are more helpful for the proposed task, which requires detecting the region of objects in the image. 
    \item When we remove the gated-fusion module, all metrics significantly drop, indicating that the absence of dynamic feature selection can lead to the model indiscriminately accepting cross-modal features in a suboptimal state.
    \item When we remove the entire selective attention component, the model performance decreases even worse due to the lack of interaction between textual and visual features. The relation and entity accuracy is higher than \textit{w/o gated}, as the interaction will not greatly impact the learning of entity and relation prediction for each modality.
    \item When we remove the query-based transformer, the model degrades substantially and is close to the pipeline baselines. The reason is that the model cannot effectively handle three sub-tasks without queries, which verifies that the module is efficient for multi-task learning.
\end{itemize}

\subsection{Modal-fusion Analysis}

This section analyzes the fusion learning of textual and visual modalities in the different modules of the proposed model. We visualize the visual cross-attention on the text and the query's cross-attention on the image to observe what kind of text tokens and image regions are selected. We chose the first sample in the test dataset directly. The detailed analyses of the modal fusion are listed as follows:

\begin{itemize}
    \item The chosen sample \textit{The Yankee Death Star is Back} requires extracting the head entity \textbf{Yankee}. Figure \ref{fig:text} shows that most of the 0-15 regions of the image (4x4 feature map) accurately focus on the token \textbf{Yankee}, which indicates that our attention and fusion mechanism can effectively help the images select the necessary head entities.
    \item Figure \ref{fig:text} shows the visual encoder pays rare attention to the special tokens such as \textit{[CLS]}, \textit{the} and \textit{[SEP]}, but more attention is allocated to other tokens that may help understand the semantics of the text. The result proves that visual features know which tokens are helpful.
    \item The query serves as the core of solving the proposed task. Table \ref{tab:table4} shows the impact of different numbers of queries on model performance. According to Table \ref{tab:table4}, the best F1 score is obtained when the number of queries is set to 5. Further, performance is drastically degraded when only 1 query is assigned because multiple outputs are impossible to handle. Moreover, the performance is also degraded when many queries are given because of the overly dispersed nature of the learned features, even though they are able to cope with multiple outputs.
    \item Using the same sample, we visualized the cross-attention map to observe query-based modal learning ability. As shown in Figure \ref{fig:query}, to match the attention map, we divided the original image into 4x4 patches, and the red box in the image indicates the object region we should detect. According to query No.1, it correctly learns the approximate location of the object, which verifies that the query-based method can understand the visual features to find the corresponding object (tail entity) based on the prior-learned textual features.
    \item We further selected \textit{query} \textit{No.3} and \textit{query} \textit{No.5} from 5 queries. these queries allocate the most attention to edges and surroundings, suggesting they learn features near the object. Even though each query is not interested in the exact same features, they can pass information to each other after the self-attention based on the transformer design and eventually extract the desired object region.
\end{itemize}

\begin{table}[ht]
    \centering
    \begin{tabular}{c | c c c c c}
    \toprule
         Metrics & 1 & 2 & \textbf{5} & 7 & 15 \\
         \hline
         F1. & 32.43 & 39.58 & \textbf{49.65} & 48.88 & 47.06 \\ 
         Rel Acc. & 44.54 & 54.45 & \textbf{68.31} & 64.91 & 56.85 \\
         Ent Acc. & 71.52 & 80.59 & \textbf{87.36} & 81.40 & 50.72 \\
    \bottomrule
    \end{tabular}
    \caption{Impact of number of queries on model performance.}
    \label{tab:table4}
\end{table}

According to the above observation, our model can understand and fuse cross-modal information, and the queries can further learn the features of different regions in the image. Therefore, these results verify that the QEOT model can efficiently predict the entity span, relation type, and object region to output the relational triples.

\section{Conclusion}

In this paper, we introduce a novel task called multimodal entity-object relational triple extraction to advance the research on the detection and semantic understanding of the case of entity pairs present in different modalities. We propose the QEOT (query-based entity-object transformer) model for this task. We designed selective attention and gated-fusion mechanisms to fully explore textual and visual features and achieve cross-modal feature fusion. To fulfill the entity extraction, relation classification, and object detection tasks simultaneously, we adopt a set of queries with cross-modal features to learn autonomously and assign the optimal matching for jointly optimizing the performance of these tasks via the Hungarian algorithm. We validated the effectiveness of our method through detailed experiments.

\bibliographystyle{plain}
\bibliography{qeot}


\appendix
\section{Appendix}
\subsection{MORTE Dataset}
As shown in Table \ref{tab:table1}, our dataset is more complex than the MORE task because it has 1681 entities, which can only be extracted accurately by understanding the text data. In addition, on average, there are 3.8 different objects and 5.75 triples for each image-text pair, which prevents us from directly adopting the previous multimodal relation extraction approach to the problem. Therefore, our dataset and task call for the capability to process and understand multimodal content.

\begin{table}[ht]
    \centering
     \caption{The MORTE dataset details statistics. Here, Img: images, Ent: entities, Triple (ave): triple facts and average number for each sample, Rel: relations}
    \begin{tabular}{c | c c c r c}
    \hline
        \textbf{MORTE} & \textbf{Img} & \textbf{ Text} & \textbf{Ent} & \textbf{Triple (ave)}& \textbf{Rel} \\
    \hline
        \textbf{Train}  & 3,037 & 3,030 & 1,270 & 17,228 (5.67) & 21 \\
        \textbf{Test} & 522  & 521  & 411  & 3,036 (5.82) & 21 \\
    \hline
        \textbf{Total} & 3,559 & 3,551 & 1,681 & 20,264 (5.75) & 21 \\
    \hline
    \end{tabular}
    \label{tab:table1}
\end{table}
\subsection{Query-based Transformer}
\begin{figure}[ht]
    \centering
    \includegraphics[width=7cm]{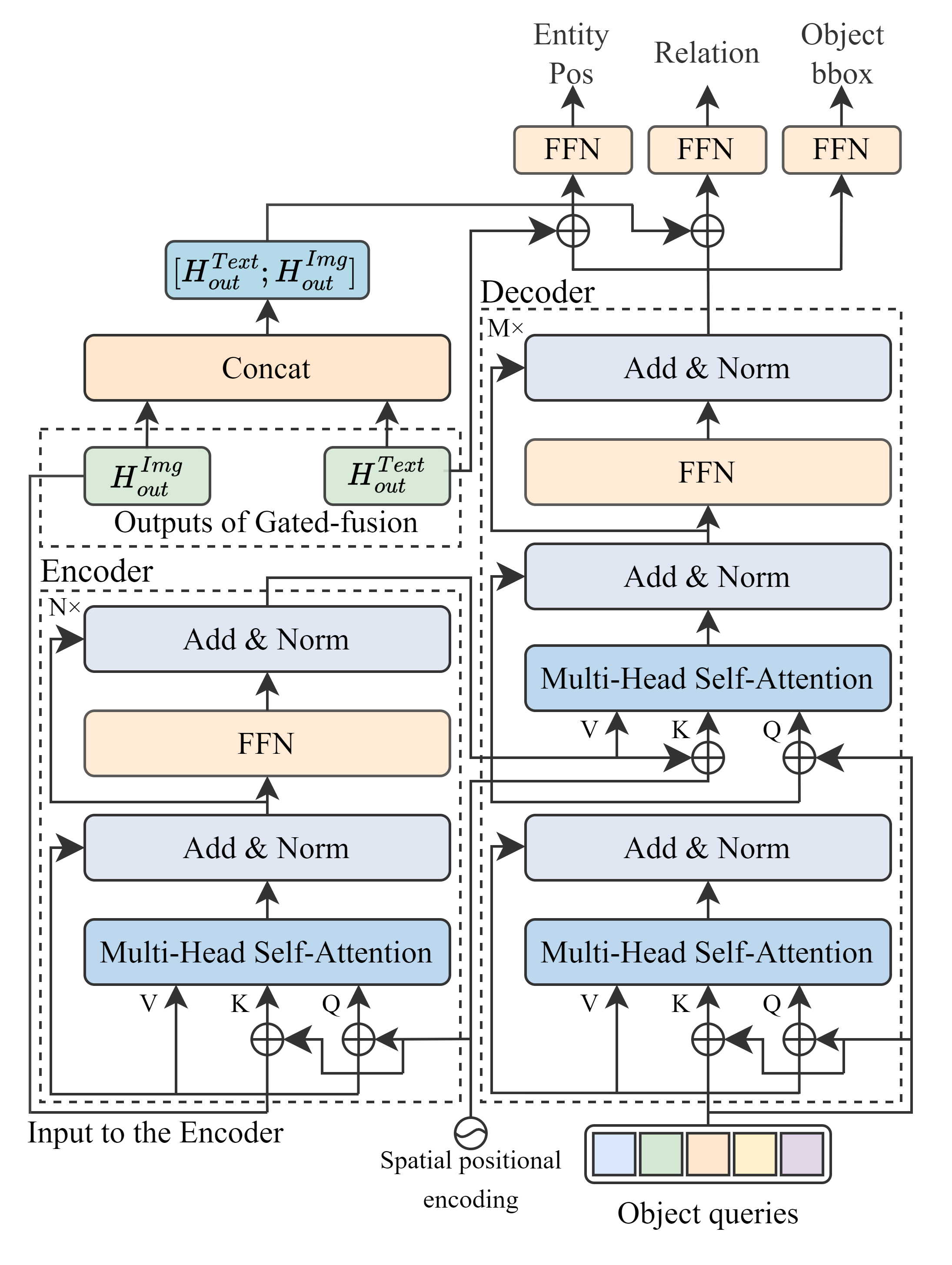}
    \caption{Detailed encoder-decoder architecture of the \\ query-based transformer.}
    \label{fig:trans}
\end{figure}

Figure \ref{fig:trans} shows the detailed query-based transformer architecture

\subsection{Entity-object Relational Triple Evaluation}
In this paper, we focus on evaluating the multimodal entity-object relational triple extraction task. An image-text pair may produce multiple triples $y = \{(e, r, b)_c\}_{c=1}^C$ in this task. The FPR performance should be satisfactory when the model understands which relation types should be paired with the entities based on the text and image.

Since it involves the object detection task, it is challenging to use FPR to evaluate triples. However, as the object in this task does not have a corresponding class but instead generates semantic associations as the tail entity with the head entity (from text), we cannot directly use mAP to evaluate it either. An intuitive solution would be to treat the head entity as the class of object, but as shown in Table \ref{tab:table1}, there are a total of 1681 different entities, which would be a massive challenge with such a dataset size. 

\textbf{To this end}, we formulate ntity-object relational triple evaluation algorithm to optimally match object positions given the entity-relation pair with ground truth and to determine the FPR evaluation of all triples. Algorithm \ref{alg:alg1} shows the pseudocode of the algorithm.

\begin{algorithm}[ht]
    \renewcommand{\algorithmicrequire}{\textbf{Input:}}
    \renewcommand{\algorithmicensure}{\textbf{Output:}}
    \caption{Entity-object relational triple evaluation algorithm}\label{alg:alg1}
    \begin{algorithmic}[1]
        \REQUIRE predicted $(E_p, R_p, B_p)$, target $(E_t, R_t, B_t)$
        \ENSURE $F1$, \, $Prec$, \, $Rec$
        \STATE $E$ for \textbf{E}ntity, $R$ for \textbf{R}elation, $B$ for \textbf{B}ounding box
        \STATE initial hash map: $\mathcal{M}_t = \varnothing$, $\mathcal{M}_p = \varnothing$, and IoU threshold: $\theta$
        \STATE initial: $tp \gets 10^{-9}$, $fp \gets 10^{-9}$, $fn \gets 10^{-9}$
        \STATE get valid index: $\mathcal{I} \gets R_p(r_p \neq \varnothing)$
        \STATE $E_p^{I} \gets E_p[\mathcal{I}]$, $R_p^{I} \gets R_p[\mathcal{I}]$, $R_p^{I} \gets R_p[\mathcal{I}]$
        \FORALL{$(e_t, r_t, b_t)$ in $(E_t, R_t, B_t)$}
            \IF{$(e_t, r_t)$ not in $\mathcal{M}_t$}
                \STATE $\mathcal{M}_t[(e_t, r_t)] = b_t$
            \ELSE
                \STATE $\mathcal{M}_t[(e_t, r_t)]$ append $b_t$
            \ENDIF
        \ENDFOR
        \STATE similarly, update: $\mathcal{M}_p \gets (E_p^{I}, R_p^{I}, B_p^{I})$
        \FORALL{$(e_t, r_t)$ in $\mathcal{M}_t$}
            \IF{$(e_t, r_t)$ not in $\mathcal{M}_p$}
                \STATE update $fn$
            \ELSE
                \STATE $b_t \gets \mathcal{M}_t[(e_t, r_t)]$, $b_p \gets \mathcal{M}_p[(e_t, r_t)]$
                \STATE cost $ \gets \Vert{b_t - b_p}\Vert_1$
                \STATE get assignment index map: $\mathcal{M}_{I} \gets$ \textit{Hungarian}(cost)
                \FORALL{$(p_i, t_i)$ in $\mathcal{M}_{I}$}
                    \STATE compute: $iou \gets IoU(b_p[p_i], b_t[t_i])$
                    \IF{$iou > \theta$}
                        \STATE update $tp$
                    \ELSE
                        \STATE update $fp$
                    \ENDIF
                \ENDFOR
            \ENDIF
        \ENDFOR
        \STATE update $F1, Prec, Rec$ with $tp, fp, fn$
        \RETURN $F1$, \, $Prec$, \, $Rec$
    \end{algorithmic}
    \end{algorithm}

\begin{figure}[ht]
    \centering
    \includegraphics[width=8cm,height=3cm]{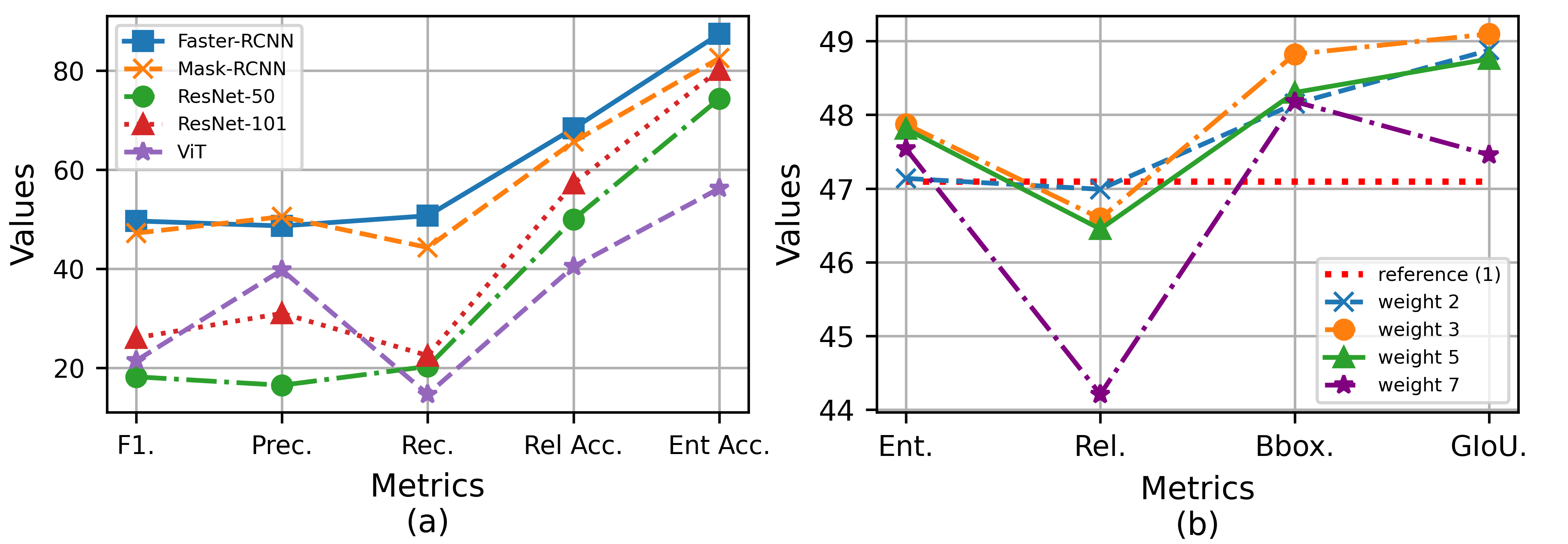}
    \caption{Impact study of different visual backbone (a) and different joint optimization weights (b).}
    \label{fig:bck}
\end{figure}

\begin{figure}[ht]
    \centering
    \includegraphics[width=8cm,height=3cm]{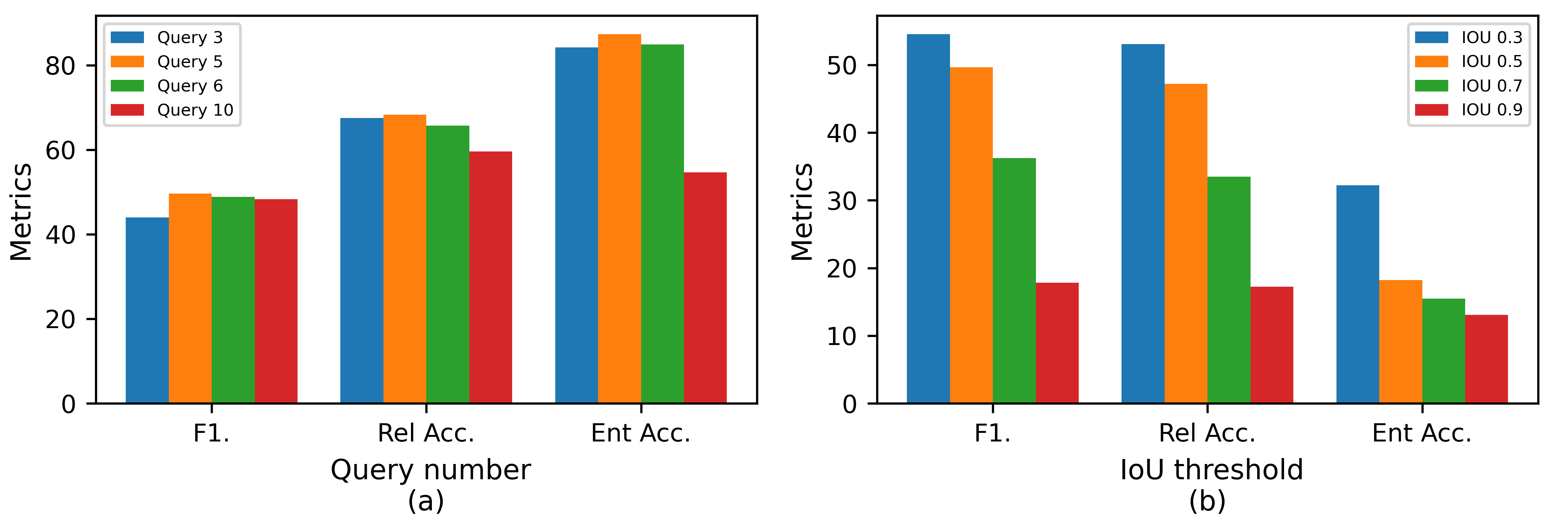}
    \caption{Impact study of different query numbers (a) and different iou threshold (b).}
    \label{fig:qiou}
\end{figure}

\subsection{Parameter Settings}
The hyperparameter settings also significantly impact model performance, and we list the main parameter settings here. Considering the convergence speed, we set the batch size to 8 and the learning rate to $3e^{-5}$. The dimension of the textual and visual embedding is 768. All models adopt the $AdamW$ as the training optimizer. For our query-based and multi-task learning model, we set the query number to 5, the encoder and decoder layers of query-based transformer are all set to 6, the iou threshold is 0.5, and the jointly optimized weights $[\lambda_{ent}, \lambda_{rel}, \lambda_{L1}, \lambda_{giou}]$ to $[1., 2., 3., 3.5]$ respectively. Additionally, we train our models on the NVIDIA A40 server with 48G GPU memory.

\subsection{Detailed Analysis}
This section mainly analyzes the effect and sensitivity of several important hyper-parameters in the query-based transformer and the joint optimization process.

Based on Figure \ref{fig:bck}a, Faster-RCNN is better than Mask-RCNN except for the Triple Precision because Mask-RCNN also takes on the semantic segmentation task, which leads to more difficult transfer learning. In contrast, the fine-grained data used for segmentation makes it more precise. In addition to the task characteristics, the small dataset size limits these models' performance, and the results of the ResNet models with the inductive bias outperform ViT, proving our guess.

\end{document}